\documentclass[12pt]{article}
\usepackage{amsmath,amsthm}
\usepackage{cite,times}
\usepackage[margin=1in,a4paper,includeheadfoot]{geometry}
\usepackage[small]{titlesec}

\numberwithin{equation}{section}

\newtheorem*{theorem}{Theorem}

\newcommand{\F}[4]{%
    {}\,{}_2F_1\biggl(%
    \genfrac{}{}{0pt}{}{#1\,,\:#2}{#3}
    \bigg\vert#4\biggr)}

\newcommand{\Ffourthree}[8]{%
    \,{}_4F_3\biggl(%
    \genfrac{}{}{0pt}{}{#1\,,\:#2\,,\:#3\,,\:#4}{#5\,,\:#6\,,\:#7}
    \bigg\vert#8\biggr)}

\titlelabel{\thetitle.\ \;}

\begin{document}
\vspace*{0pt}

\begin{center}
\Large\textbf{
On the refined $3$-enumeration of alternating sign matrices
}
\end{center}

\begin{center}
\large{
F. Colomo\textsuperscript{a)}
and A. G. Pronko\textsuperscript{b)}
}
\end{center}

\begin{center}
\small\textsl{
\textsuperscript{a)} I.N.F.N., Sezione di Firenze,
and Dipartimento di Fisica, Universit\`a di Firenze,\\
Via G. Sansone 1, 50019 Sesto Fiorentino (FI), Italy\\
\textsuperscript{b)} Sankt-Petersburg Department of V. A. Steklov
Mathematical Institute of Russian\\ Academy of Sciences,
Fontanka 27, 191023 Sankt-Petersburg, Russia
}
\end{center}

\begin{abstract}
An explicit expression for the numbers $A(n,r;3)$
describing the refined $3$-enumeration of
alternating sign matrices is given.
The derivation is based on the recent
results of Stroganov for the corresponding generating function.
As a result, $A(n,r;3)$'s are represented as $1$-fold
sums which can also be written in terms of
terminating ${}_4F_3$ series of argument $1/4$.
\end{abstract}

\section{Introduction}

An alternating sign matrix (ASM) is a matrix of $1$'s, $0$'s and $-1$'s
such that in each row and in each column the first and the last
nonzero entry is 1, and all  nonzero entries
alternate  in sign. There are many nice results concerning ASMs
(for a review, see \cite{Br-99}). The most celebrated result
gives the total number $A(n)$ of $n\times n$ ASMs.
It was conjectured by Robbins, Mills and Rumsey \cite{MRR-82,MRR-83}
and subsequently proved by Zeilberger \cite{Z-96a} and Kuperberg
\cite{Ku-96} that
\begin{equation}\label{An}
A(n)=\prod_{k=1}^{n}\frac{(3k-2)!\, (k-1)!}{(2k-1)!\,
(2k-2)!} =
\prod_{k=1}^{n}   \frac{(3k-2)!}{(2n-k)!}.
\end{equation}
Yet another formula was conjectured in \cite{MRR-82,MRR-83},
and proved in \cite{Z-96b}, concerning the refined enumeration
of ASMs, namely, that the number of $n\times n$ ASMs
with their sole $1$ of the first (or last) column (or row)
at the $r$-th entry, denoted as $A(n,r)$, is
\begin{equation}\label{Anr}
A(n,r)
=\frac{\binom{n+r-2}{n-1}\binom{2n-1-r}{n-1}}{\binom{3n-2}{n-1}}\;
A(n)\;.
\end{equation}
One more conjecture of Robbins, Mills and Rumsey, proved by
Kuperberg \cite{Ku-96}, gives the number of $3$-enumerated ASMs.
In general, $x$-enumeration counts
ASMs with a weight $x^k$, where $k$ is the total number of
$-1$'s in each matrix (the number $x$ here should not be
confused with the variable $x$ below).
The total number of $3$-enumerated ASMs
is denoted as $A(n;3)$ and the result reads
\begin{equation}\label{An3}
A(2m+1;3)= 3^{m(m+1)}
\prod_{k=1}^{m} \left[\frac{(3k-1)!}{(m+k)!}\right]^2
\qquad
A(2m+2;3)= 3^m \frac{(3m+2)!\, m!}{\bigl[(2m+1)!\bigr]^2}\; A(2m+1;3)\;.
\end{equation}

In the present paper we study the problem of the
refined $3$-enumeration of ASMs, i.e., we derive
the numbers $A(n,r;3)$. The main result can be summarized as follows.
\begin{theorem}[Main result]
The refined 3-enumeration of ASMs, $A(n,r;3)$, is given by
\begin{equation}\label{theor1}
\begin{split}
A(2m+2,r;3)&=\frac{b(m,r-1)+b(m,r-2)}{2}\;A(2m+2;3)
\\
A(2m+3,r;3)&=\frac{2\,b(m,r-1)+5\,b(m,r-2)+2\,b(m,r-3)}{9}\;A(2m+3;3)
\end{split}
\end{equation}
where the quantities $b(m,\alpha)$,
obeying $b(m,\alpha)=b(m,2m-\alpha)$,
are given by
\begin{multline} \label{theor2}
b(m,\alpha)=\frac{(2m+1)!\,m!}{3^m\,(3m+2)!}
\sum_{\ell=\max(0,\alpha-m)}^{[\alpha/2]}
(2m+2-\alpha+2\ell)\binom{3m+3}{\alpha-2\ell}
\\ \times
\binom{2m+\ell-\alpha+1}{m+1} \binom{m+\ell+1}{m+1}2^{\alpha-2\ell}
\end{multline}
for $\alpha=0,1,\dots,2m$, while they are
assumed to be zero for all other values of $\alpha$.
\end{theorem}

As a comment to the main result it is to be mentioned
that the numbers $A(n,r;3)$  appear
not to be writable  as a single hypergeometric term, i.e.,
not to be `round' (or `smooth'), contrarily to the  expressions
for $A(n)$, $A(n,r)$ and $A(n;3)$.
For instance, direct inspection of the results
of computer experiments for $A(n,r;3)$ shows
the appearance of large prime
factors in their prime factorizations, strongly
suggesting  that no answer
for $A(n,r;3)$ can be given in the form of a single hypergeometric
term.  This explains why no conjecture concerning $A(n,r;3)$ was given
previously, and it might moreover imply that any closed expression for
$A(n,r;3)$ could be at best a  sum of several `round' terms. In
view of this, the expression in formulae \eqref{theor1} and
\eqref{theor2},
even if not as elegant and neat as other results
in the context of ASM enumerations, is probably
the best that can be achieved: a $1$-fold sum of `round' terms.

Our proof of the Theorem is based on a certain representation
for the generating function for the numbers $A(n,r;3)$, recently
obtained by Stroganov \cite{S-03}.
In the next Section we recall the relevant formulae
of paper \cite{S-03} and shortly review their backgrounds.
In Section 3 we apply standard relations
for Gauss hypergeometric function to simplify significantly
Stroganov's expression  for the generating function of
$3$-enumerated ASMs. In Section 4, from
this simplified expression we obtain
explicit formulae for $A(n,r;3)$,
thus proving the Theorem.

\section{Preliminaries}

In studying the refined $3$-enumeration of $n\times n$ ASMs
it is convenient to define the generating function
\begin{equation}\label{defH}
H_n^{(3)} (t)
=\frac{1}{A(n;3)}\sum_{r=1}^{n} A(n,r;3)\, t^{r-1}\;,\qquad
H_n^{(3)}(1)=1\;.
\end{equation}
A simple property of the generating function is
$H_n^{(3)}(t)=t^{n-1} H_n^{(3)}(t^{-1})$, which follows
from the relation $A(n,r;3)=A(n,n-r+1;3)$ expressing
left-right (top-bottom) symmetry within the set of
$n\times n$ ASMs.

In paper \cite{S-03} the following defining
properties of the generating function have been established.
The generating functions $H_n^{(3)} (t)$
for $n$ even, $n=2m+2$, and for $n$ odd, $n=2m+3$,
are related by the formulae
\begin{equation}\label{even-odd}
H_{2m+2}^{(3)}(t)=\frac{(t+1)}{2}\; B_{2m}(t)\;,\qquad
H_{2m+3}^{(3)}(t)=\frac{(2t+1)(t+2)}{9}\; B_{2m}(t)\;,
\end{equation}
where $B_{2m}(t)$ is some polynomial of degree $2m$.
It is convenient to define
$E_m(t):=t^{-m} B_{2m}(t)$ so that $E_m(t)=E_m(t^{-1})$.
The function $E_m(t)$ can be related to another function
$V_m(x)$, obeying, in turn, the relation $V_m(x)=V_m(x^{-1})$, by
means of the transformation
\begin{equation}\label{Et}
E_m(t)=\frac{V_m(x)}{(x-1+x^{-1})^{m}}\;,\qquad
t=-\frac{x-q}{qx-1}
\end{equation}
or, inversely,
\begin{equation}\label{Vx}
V_m(x)= \frac{3^m E_m(t)}{(t+1+t^{-1})^m}\;,\qquad
x=\frac{t+q}{tq+1}
\end{equation}
where (and everywhere below) the following shorthand notation is used
\begin{equation}
q:=\exp(\mathrm{i}\pi/3)\;.
\end{equation}
The function
\begin{equation}\label{FviaV}
h_m(x):=(x-x^{-1})^{2m+1} (x+2+x^{-1})\, V_m(x)
\end{equation}
is found to be
\begin{equation}\label{Fm-fg}
h_m(x) = c_m \biggl(g_m(x) + \frac{3m+2}{3m+1}\, f_m(x) \biggr)
\end{equation}
where the functions $f_m(x)$ and $g_m(x)$ are given by
\begin{align}
g_m(x)
&:= \sum_{k=0}^{m} \binom{m+2/3}{k}\binom{m-2/3}{m-k}
\bigl(x^{3m+2-6k}-x^{-3m-2+6k}\bigr)
\\ \label{fg}
f_m(x)
&:= \sum_{k=0}^{m} \binom{m+1/3}{k}\binom{m-1/3}{m-k}
\bigl(x^{3m+1-6k}-x^{-3m-1+6k}\bigr)
\end{align}
and $c_m$ is some constant such that $V_m(1)=E_m(1)=1$.
For the reader's convenience we note that the variable
$x$ here and the variable $u$ of Ref.~\cite{S-03}
are related by $\mathrm{e}^{2\mathrm{i}u}=-x^{-1}$.

The result just described for the generating function of
$3$-enumerated ASMs was obtained by analyzing
Izergin's determinant formula \cite{I-87} for the partition function
of the six-vertex model with domain wall boundary conditions
(also known as `square ice').
The one-to-one correspondence of square ice states with ASMs was
pointed out in \cite{EKLP-92} (see also \cite{RR-86})
and fruitfully applied by Kuperberg in \cite{Ku-96}.
The approach proposed in \cite{S-02} for $1$-enumerated ASMs
and extended to $3$-enumerated ASMs in \cite{S-03},
uses a certain functional equation for the
square ice partition function.
This functional equation is nothing but
the Baxter's T-Q equation for the ground state of
XXZ Heisenberg spin-$1/2$ chain at $\Delta=-1/2$ and with odd
number of sites $N=2m+1$, studied previously in \cite{S-01}.

In particular, expression \eqref{Fm-fg} for $h_m(x)$
arises as the solution of the functional equation
\begin{equation}\label{TQ}
y(x)+ y(q^2x) + y(q^{-2}x)=0
\end{equation}
which is supplemented by the condition that
its solution $y(x)=h_m(x)$ must have the structure \eqref{FviaV}
where the function $V_m(x)$ is supposed to be such that
$V_m(\mathrm{e}^{\mathrm{i}\varphi})$
is an even trigonometric
polynomial of degree $m$ in the variable $\varphi$.
This condition allows one to reduce the functional equation
to a finite set of linear algebraic equations. These equations
can be solved explicitly thus leading to the formula for $h_m(x)$.

The functions $f_m(x)$ and $g_m(x)$ are also solutions
of equation \eqref{TQ}, and by construction they
have the structure
\begin{equation}\label{fQgP}
f_m(x)=(x-x^{-1})^{2m+1} Q_m(x)\;,\qquad
g_m(x)=(x-x^{-1})^{2m+1} P_{m+1}(x),
\end{equation}
where $Q_m(x)$ and $P_{m+1}(x)$ are such that
$Q_m(\mathrm{e}^{\mathrm{i}\varphi})$ and
$P_{m+1}(\mathrm{e}^{\mathrm{i}\varphi})$
are even trigonometric polynomials of degrees $m$ and $m+1$
respectively. These two functions are two
independent solutions of T-Q equation.
Note that as the second independent solution (i.e., that solution which
is of higher degree) one can also regard the function
$P_{m+1}(x)+\frac{3m+2}{3m+1}\; Q_m(x)$ as well as any function of
the form $P_{m+1}(x)+\alpha\, Q_m(x)$, see \cite{PS-99}.
Thus, \eqref{Fm-fg} just implies
that $3$-enumerated ASMs are described by the second independent
solution of T-Q equation.  The first solution, $Q_m(x)$, was shown in
\cite{S-02} to be related to $1$-enumerated ASMs.

Namely, in the case of $1$-enumerated ASMs it was shown that
\begin{equation}\label{HQ}
H_n^{(1)} (t)= \text{const}\times
\frac{Q_{n-1}(x)}{(qx-q^{-1}x^{-1})^{n-1}}\;,\qquad
t=\frac{qx^{-1}-q^{-1}x}{qx-q^{-1}x^{-1}}\;.
\end{equation}
where the generating function
$H_n^{(1)} (t)$ is defined similarly to \eqref{defH} by
\begin{equation}\label{gen-1}
H_n^{(1)} (t)
:=\frac{1}{A(n)}\sum_{r=1}^{n} A(n,r)\, t^{r-1}\;,\qquad
H_n^{(1)}(1)=1\;.
\end{equation}
To reproduce formulae \eqref{An} and \eqref{Anr}
from \eqref{HQ} it was proposed
to use the differential equation satisfied by $f_m(x)$, which reads
\begin{equation}\label{eq-f}
\left[\partial_\varphi^2
-6m \cot3\varphi\;\partial_\varphi
-(3m+1)(3m-1)\right]
f_m(\mathrm{e}^{\mathrm{i}\varphi})=0\;.
\end{equation}
Fortunately, the corresponding equation
on the generating function appears to be
a hypergeometric equation
\begin{equation}
\bigl[(t-1)t\,\partial_t^2 +2 (t+n-1)\,\partial_t-n(n-1)\bigr]
H_n^{(1)}(t)=0
\end{equation}
and its polynomial solution gives
\begin{equation}\label{Hfin}
H_n^{(1)}(t) = \frac{(2n-1)!\,(2n-2)!}{(3n-2)!\,(n-1)!}\;
\F{-n+1}{n}{-2n+2}{t}.
\end{equation}
Here the proper normalization, see \eqref{gen-1}, follows from
the Chu-Vandermonde identity
\begin{equation}\label{Gauss-sum}
\F{-m}{b}{c}{1}
=\frac{(c-b)_m}{(c)_m};\qquad
(a)_m:=a(a+1)\cdots(a+m-1).
\end{equation}
Since $A(n,1)=A(n-1)$, Eqn.~\eqref{Hfin} results in an elementary
recursion
\begin{equation}
\frac{A(n-1)}{A(n)}=H_n^{(1)}(0)=
\frac{(2n-1)!\,(2n-2)!}{(3n-2)!\,(n-1)!}
\end{equation}
which, supplemented by the initial condition $A(1)=1$, gives
formula \eqref{An}.
Expanding RHS of \eqref{Hfin} in power
series in $t$ and multiplying these coefficients by  $A(n)$,
one reproduces, in turn, formula \eqref{Anr}.

Unfortunately, a similar procedure cannot be applied in the case
of $3$-enumerated ASMs.  It can be verified directly that the
function $h_m(x)$ solves the equation
\begin{equation}\label{eq-h}
\biggl\{\partial_\varphi^2
-3\biggl[(2m+1)\cot3\varphi
-\frac{1}{\sin3\varphi}\biggr]\partial_\varphi
-(3m+1)(3m+2)\biggr\}h_m(\mathrm{e}^{\mathrm{i}\varphi})=0.
\end{equation}
This equation leads to a rather bulky
non-hypergeometric equation
for the generating function $H_n^{(3)}(t)$, which
can hardly be solved by standard means.
Moreover, even though the function $Q_m(x)$ can easily be deduced
from the equation for $f_m(x)$ above, the differential equation
for $g_m(x)$ (for this equation, see \cite{S-01})
does not allow one to find the function $P_{m+1}(x)$ in a similar
way. As a matter of fact,
only an implicit answer for $H_n^{(3)}(t)$ in terms of two recurrences,
generated by three-terms recurrences for $f_m(x)$ and $g_m(x)$, was given
in paper \cite{S-03}.
Here we would like just to mention that
from differential equation \eqref{eq-h} it can easily be seen  that the
operator $(\sin3\varphi)^{-1}\partial_\varphi$
plays the role of a `lowering' operator in the set of trigonometric
polynomials $\{h_m(\mathrm{e}^{\mathrm{i}\varphi})\}$. It is thus
straightforward to rewrite Eqn.~\eqref{eq-h}
as a three term recurrence for $h_m(x)$ and, hence, to obtain a single
recurrence for $H_n^{(3)}(t)$. However this point of view will not be
developed  further, since it amounts merely to a
reformulation of the problem. Instead, in the next Section we
show how to overcome
the difficulty and  find the function $H_n^{(3)}(t)$ explicitly.

As a last comment here we would like to mention
that, although the previous results on the refined $3$-enumeration
of ASMs run over papers
\cite{S-03,S-02,S-01}, the
present research have originated from
our study of the square ice boundary correlators obtained
in \cite{BPZ-02}. Within our study
the refined enumerations of ASMs
arise as solutions of some differential equations
rather than functional ones, in particular,
the expression for $h_m(x)$
arises as a solution of \eqref{eq-h}.
The illustration of this approach is out of the scope
of the present paper and will be given elsewhere \cite{CP-04};
here we restrict ourselves to obtaining the numbers $A(n,r;3)$
from the known $h_m(x)$.

\section{The generating function}

We begin with noticing that the functions $f_m(x)$ and $g_m(x)$
can be also written as follows
\begin{align}\label{gFF}
g_m(x)&=\frac{\Gamma(m+1/3)}{m!\,\Gamma(1/3)}
\biggl[
x^{3m+2}\F{-m}{-m-2/3}{1/3}{x^{-6}}
\notag\\ & \mspace{212mu}
- x^{-3m-2}\F{-m}{-m-2/3}{1/3}{x^{6}}\biggr],
\\
\label{fFF}
f_m(x)&=\frac{\Gamma(m+4/3)}{m!\,\Gamma(4/3)}
\biggl[
x^{-3m+1}\F{-m}{-m+1/3}{4/3}{x^{6}}
\notag\\  & \mspace{212mu}
- x^{3m-1}\F{-m}{-m+1/3}{4/3}{x^{-6}}
\biggr].
\end{align}
Since the parameters of the hypergeometric functions
entering these expressions
differ by integers one can expect
that $g_m(x)$ and $f_m(x)$ are connected by some three-term
relations via Gauss relations (see, e.g., Sect.~2.8 of \cite{E-81}).
Indeed, using Gauss relations it can be shown that
\begin{multline}
\F{-m}{-m-2/3}{1/3}{z}
=\frac{3m+4}{2}\;\F{-m-1}{-m-2/3}{4/3}{z}
\\
-\frac{3m+2}{2}\;(1+z)\F{-m}{-m+1/3}{4/3}{z}
\end{multline}
and therefore we have the following relation
\begin{equation}\label{gff}
g_m(x)=\frac{3m+2}{2(3m+1)}\;
(x^3+x^{-3})\,f_m(x) -
\frac{3(m+1)}{2(3m+1)}\; f_{m+1}(x).
\end{equation}
Substituting \eqref{gff} into \eqref{Fm-fg} we obtain
\begin{equation}\label{hff}
h_m(x)=c_m\,\frac{3m+2}{2(3m+1)}
\left[
(x^3+2+x^{-3})\, f_m(x) - \frac{3m+3}{3m+2}\; f_{m+1}(x)
\right].
\end{equation}
Comparing Eqns.~\eqref{gff} and \eqref{hff} with Eqns.~\eqref{FviaV}
and \eqref{fQgP}
we find that
\begin{equation}\label{PQQ}
P_{m+1}(x)=
\frac{3m+2}{2(3m+1)}\;
(x^3+x^{-3})\,Q_m(x) -
\frac{3(m+1)}{2(3m+1)}\; (x-x^{-1})^2\,Q_{m+1}(x)
\end{equation}
and
\begin{equation}\label{VQQ}
V_m(x)= c_m\,\frac{3m+2}{2(3m+1)}
\left[(x-1+x^{-1})^2\,Q_m(x)- \frac{3m+3}{3m+2}\,
(x-2+x^{-1})\, Q_{m+1}(x)
\right].
\end{equation}
Hence all functions in question are expressed in terms of $Q_m(x)$'s.

The advantage of this approach is based on the fact that
for the function $Q_m(x)$ the following explicit formula
can be found
\begin{equation}\label{QF}
Q_m(x)
=\frac{(2m)!}{3^m\, (m!)^2}\;
\biggl(\frac{qx^{-1}-q^{-1}x}{q-q^{-1}}\biggr)^m
\F{-m}{m+1}{-2m}{\frac{q x-q^{-1} x^{-1}}{qx^{-1}-q^{-1}x}}.
\end{equation}
As already mentioned in our preliminary comments,
the function $Q_m(x)$ can be deduced  from the
differential equation for $f_m(x)$ above (the proper
normalization, for instance,  can be found using the  three-term
relation for $f_m(x)$'s).
Here we give a proof of \eqref{QF} by a straightforward
transformation of the function $f_m(x)$ to the form
$f_m(x)=(x-x^{-1})^{2m+1}Q_m(x)$.

The key identity which is to be used here is
the so-called cubic transformation of Gauss hypergeometric
function \cite{E-81} which in its most symmetric form reads:
\begin{multline}\label{cubic}
\frac{\Gamma(a)}{\Gamma(2/3)}\;
\F{a+1/3}{a}{2/3}{z^3} - \omega^{-1} z\,
\frac{\Gamma(a+2/3)}{\Gamma(4/3)}\;
\F{a+1/3}{a+2/3}{4/3}{z^3}
\\
= 3^{-3a+1}{\Bigl(\frac{1-z}{1-\omega}\Bigr)^{-3a}}
\frac{\Gamma(3a)}{\Gamma(2a+2/3)}\;
\F{a+1/3}{3a}{2a+2/3}{\omega\frac{z-\omega}{1-z}}.
\end{multline}
Here $\omega$ is a primitive cubic root of unity,
$\omega=\exp(\pm2\mathrm{i}\pi/3)$, and $a$ is arbitrary parameter.
To show that indeed the cubic transformation is relevant to
our case, let us rewrite \eqref{fFF} in the form
consistent with LHS of \eqref{cubic}.
Taking into account that
\begin{equation}
\F{-m}{-m+1/3}{4/3}{z} =\frac{\Gamma(1/3)\,\Gamma(4/3)}
{\Gamma(-m+1/3)\,\Gamma(m+4/3)} (-z)^m
\F{-m}{-m-1/3}{2/3}{z^{-1}}
\end{equation}
and
\begin{equation}
\frac{\Gamma(1/3)}{\Gamma(m+4/3)}
=(-1)^{m+1} \frac{\Gamma(-m-1/3)}{\Gamma(2/3)}\;,\qquad
\frac{\Gamma(m+4/3)}{\Gamma(-m+1/3)}
=(-1)^m \frac{(3m+1)!}{3^{3m+1}\,m!}
\end{equation}
it is easy to see that \eqref{fFF} can be rewritten in the form
\begin{multline}\label{f-ready}
f_m(x)=\frac{(-1)^{m+1}(3m+1)!}{3^{3m+1}\, (m!)^2}\;
x^{3m+1}\biggl[
\frac{\Gamma(-m-1/3)}{\Gamma(2/3)}\;
\F{-m}{-m-1/3}{2/3}{x^{-6}}
\\
+x^{-2} \frac{\Gamma(-m+1/3)}{\Gamma(4/3)}\;
\F{-m}{-m+1/3}{4/3}{x^{-6}}
\biggr].
\end{multline}
Clearly, both terms in the brackets are the same as in LHS of
\eqref{cubic} provided the parameter $a$ is specialized to the
value $a=-m-1/3$.

To apply the cubic transformation to \eqref{f-ready}
we first define
\begin{equation}\label{waz-in}
W(a;z) :=
\frac{\Gamma(a)}{\Gamma(2/3)}\;
\F{a+1/3}{a}{2/3}{z^3} + z\,
\frac{\Gamma(a+2/3)}{\Gamma(4/3)}\;
\F{a+1/3}{a+2/3}{4/3}{z^3}
\end{equation}
so that
\begin{equation}\label{fW}
f_m(x)=\frac{(-1)^{m+1}(3m+1)!}{3^{3m+1}\, (m!)^2}\;
x^{3m+1} W(-m-1/3;x^{-2}).
\end{equation}
Next, we note that for a sum of two terms one can always write
\begin{equation}
X+Y =\frac{q}{q-q^{-1}}\;\bigl(X-q^{-2}Y\bigr)-
\frac{q^{-1}}{q-q^{-1}}\bigl(X-q^2Y\bigr)
\end{equation}
and if $q=\exp(\mathrm{i}\pi/3)$, which is exactly the case,
one can set $\omega=q^2$
for the first pair of terms and $\omega=q^{-2}$ for the second one.
This recipe allows one to apply the cubic transformation,
that gives
\begin{multline} \label{waz-out}
W(a;z)=
\frac{3^{-3a+1}\,\Gamma(3a)}{\Gamma(2a+2/3)}\;
\frac{(1-z)^{-3a}}{q(1-q^2)^{-3a+1}}
\biggl[
\F{a+1/3}{3a}{2a+2/3}{q^2\frac{z-q^2}{1-z}}
\\
+q^{3a+1}
\F{a+1/3}{3a}{2a+2/3}{q^{-2}\frac{z-q^{-2}}{1-z}}
\biggr].
\end{multline}
To obtain a new formula for $f_m(x)$ via \eqref{fW} we have
to evaluate now the limit $a\to-m-1/3$ of \eqref{waz-out}.
The limit of the pre-factor can be easily found due to
\begin{equation}
\lim_{a\to-m-1/3}\frac{\Gamma(3a)}{\Gamma(2a+2/3)}
=\frac{2}{3}\;\frac{(-1)^{m+1}(2m)!}{(3m+1)!}\;.
\end{equation}
To find the limit of the expression in the brackets in \eqref{waz-out}
we note that
\begin{equation}
q^2\frac{z-q^2}{1-z}=:s\;\qquad
q^{-2}\frac{z-q^{-2}}{1-z}=1-s
\end{equation}
and hence the following formula can be used
\begin{equation}\label{limfin}
\lim_{a\to-m-\frac{1}{3}}\biggl[
\F{a+\frac{1}{3}}{3a}{2a+\frac{2}{3}}{s}
+q^{3a+1}
\F{a+\frac{1}{3}}{3a}{2a+\frac{2}{3}}{1-s}
\biggr]
=\frac{3}{2}\, \F{-m}{-3m-1}{-2m}{s}.
\end{equation}
Formula \eqref{limfin} can be proved, for instance,
by virtue of standard analytic continuation formulae for the
hypergeometric function
(see, e.g., Eqns.~(1) and (2) in \S 2.10 of \cite{E-81}).
Collecting formulae we arrive to the expression
\begin{equation}
W(-m-1/3;z)=- \frac{3^{2m+1}\,(2m)!}{(3m+1)!}
\frac{(1-z)^{3m+1}}{(q-q^{-1})^{m}}
\F{-m}{-3m-1}{-2m}{q^2\frac{z-q^2}{1-z}}.
\end{equation}
Finally, substituting this expression into \eqref{fW}
and using the identity
\begin{equation}
\F{a}{b}{c}{z} =(1-z)^{-a} \F{a}{c-b}{c}{\frac{z}{z-1}}
\end{equation}
we obtain
\begin{equation}
f_m(x)=
\frac{(2m)!}{3^m\,(m!)^2}\;
(x-x^{-1})^{2m+1}
\biggl(\frac{qx^{-1}-q^{-1}x}{q-q^{-1}}\biggr)^m
\F{-m}{m+1}{-2m}{\frac{q x-q^{-1} x^{-1}}{qx^{-1}-q^{-1}x}}.
\end{equation}
Obviously, this expression leads directly to \eqref{QF}
which is thus proved.

Hence, we have just shown that a  particular solution of
equation \eqref{TQ}, the function $f_m(x)$, Eqn.~\eqref{fg},
is connected to the `first' solution of the Baxter T-Q equation,
$Q_m(x)$, Eqn.~\eqref{QF}, via the cubic transformation.
Formulae \eqref{gff} and \eqref{hff}
mean that the same transformation is responsible for relationship
of $g_m(x)$ and $h_m(x)$ with the second independent solution of
T-Q equation, which can be chosen either as $P_{m+1}(x)$,
in the case of $g_m(x)$, or
as $P_{m+1}(x)+\frac{3m+2}{3m+1}\;Q_m(x)=(x+2+x^{-1})\,V_m(x)$,
in the case of $h_m(x)$.

To write down the resulting expressions for functions $P_{m+1}(x)$ and
$V_m(x)$, and for the purpose of subsequent transformation of
these expressions, we define
\begin{equation}\label{phi}
\Phi_m^{(k)} (x) :=
\biggl(\frac{qx^{-1}-q^{-1}x}{q-q^{-1}}\biggr)^m
\F{-m}{k+1}{-m-k}{\frac{q x-q^{-1} x^{-1}}{qx^{-1}-q^{-1}x}}.
\end{equation}
Note that for positive
integer $m$,  function $\Phi_m^{(k)}(x)$ possesses the property
$\Phi_m^{(k)}(x)=\Phi_m^{(k)}(x^{-1})$ and thus, obviously,
$\Phi_m^{(k)}(x)$
is a polynomial of degree $m$ in the variable $u:=x+x^{-1}$.

In particular, Eqn.~\eqref{QF} now reads
\begin{equation}
Q_m(x)=\frac{(2m)!}{3^m\, (m!)^2}\, \Phi_m^{(m)}(x).
\end{equation}
Formulae \eqref{PQQ} and \eqref{VQQ} give
\begin{align} \label{Ppp}
P_{m+1}(x)&=\frac{3m+2}{2(3m+1)} \frac{(2m)!}{3^m\,(m!)^2}
\biggl[ (x^3+x^{-3})\, \Phi_m^{(m)}(x)
\notag\\ &\mspace{200mu}
-\frac{2(2m+1)}{3m+2}\,(x-x^{-1})^2\, \Phi_{m+1}^{(m+1)}(x)
\biggr],
\\ \label{Vpp}
V_m(x)
&=\frac{(2m)!\,(2m+1)!}{m!\, (3m+1)!}
\biggl[(x-1+x^{-1})^2\, \Phi_m^{(m)}(x)
\notag\\ &\mspace{200mu}
-\frac{2(2m+1)}{3m+2}\, (x-2+x^{-1})\, \Phi_{m+1}^{(m+1)}(x)
\biggr].
\end{align}
Here the expression for $V_m(x)$ is written according to
the proper normalization of this function,
$V_m(1)=1$. The normalization
can be verified by virtue of Chu-Vandermonde identity
\eqref{Gauss-sum} and it corresponds to the choice
\begin{equation}
c_m=(3m+1)\,\frac{3^{m+1} m!\,(2m+2)!}{(3m+3)!}\;.
\end{equation}

Formulae \eqref{Ppp} and \eqref{Vpp} for functions $P_{m+1}(x)$
and $V_{m}(x)$ are not the final expressions for them yet
since they can be notably simplified. Indeed, again
using Gauss relations one can prove, for instance, the following
relations for the functions \eqref{phi}
\begin{align}\label{Pdec}
\Phi_{m+1}^{(k)}(x) &= (x+x^{-1})\,\Phi_{m}^{(k)}(x)
-\frac{m(m+2k+1)}{3(m+k+1)(m+k)}\,(x^2+1+x^{-2})\,\Phi_{m-1}^{(k)}(x)
\\ \label{Pmix}
\Phi_{m}^{(k+1)}(x) &= \frac{m+2k+2}{2(m+k+1)}\,\Phi_m^{(k)}(x)
+\frac{m}{2(m+k+1)}\,(x+x^{-1})\,\Phi_{m-1}^{(k+1)}(x)\;.
\end{align}
These two relations can be regarded as basic relations; many other
tree-term relations connecting $\Phi_{m}^{(k)}(x)$'s
can be derived from \eqref{Pdec} and \eqref{Pmix}. Using these
three-term relations we find that for the  function
$P_{m+1}(x)$, in particular, the following formula is valid
\begin{equation}\label{Pfinal}
P_{m+1}(x)= \frac{(2m)!}{3^m\, m! (m+1)!}
\Bigl[(3m+2)\,\Phi_{m+1}^{(m-1)}(x)-
(2m+1)\,\Phi_{m+1}^{(m)}(x)\Bigr].
\end{equation}
Similarly, for the function $V_m(x)$ we obtain
\begin{equation}\label{Vfinal}
V_m(x)= \frac{(2m)!\,(2m+2)!}{(m+1)!\,(3m+2)!}
\Bigl[(2m+1)\,\Phi_{m}^{(m+1)}(x)
-m(x-1+x^{-1})\,\Phi_{m-1}^{(m+1)}(x)\Bigr].
\end{equation}
This formula for $V_m(x)$
can be further simplified, due to \eqref{Pmix}, to be
written solely in terms of $\Phi_m{(k)}$'s, but
\eqref{Vfinal} is more convenient
for the purpose of obtaining the function
$E_m(t)$, which is of primary interest for $3$-enumerated ASMs.
Applying \eqref{Vx} we obtain
\begin{multline}\label{Efinal}
E_m(t)=
\frac{(2m)!\,(2m+2)!}{3^m\,(m+1)!\,(3m+2)!}
\biggl[(2m+1)(t+2)^m\,
\F{-m}{m+2}{-2m-1}{\frac{1+2t}{t(t+2)}}
\\
-3m\,(t+2)^{m-1} \F{-m+1}{m+2}{-2m}{\frac{1+2t}{t(t+2)}}\biggr].
\end{multline}
Formulae \eqref{Pfinal}, \eqref{Vfinal} and, especially, \eqref{Efinal}
are the main results of this Section.

\section{The numbers $A(n,r;3)$}

Let us begin with showing
how Eqn.~\eqref{An3} for $A(n;3)$ is recovered from the
just obtained expression for the generating function.
Here the property $A(n,1;3)=A(n-1;3)$ is to be used.
It implies
\begin{equation}
\frac{A(2m+1;3)}{A(2m+2;3)}
=H_{2m+2}^{(3)}(0)=\frac{1}{2}\; B_{2m}(0),\qquad
\frac{A(2m+2;3)}{A(2m+3;3)}
=H_{2m+3}^{(3)}(0)=\frac{2}{9}\; B_{2m}(0).
\end{equation}
Hence, one has the recurrences
\begin{align} \label{Aodd}
A(2m+3;3)&=\frac{9}{\bigl[B_{2m}(0)\bigr]^2}\; A(2m+1;3), &A(1;3)=1;
\\ \label{Aeven}
A(2m+2;3)&=\frac{9}{B_{2m}(0)B_{2m-2}(0)} A(2m;3), &A(2;3)=2.
\end{align}
Recall that $B_{2m}(t)=t^m E_m(t)$ in \eqref{even-odd}.
{}From \eqref{Efinal} one finds
\begin{equation} \label{Bzero}
B_{2m}(0)=
\frac{(2m+1)!\,(2m+2)!}{3^m\,(m+1)!\,(3m+2)!}.
\end{equation}
Substituting \eqref{Bzero}  into \eqref{Aodd} and \eqref{Aeven}, and
solving the recurrences one obtains \eqref{An3}.

Let us now turn to our main target, $A(n,r;3)$, which describes
the refined $3$-enumeration of ASMs. Formulae \eqref{even-odd}
imply that
\begin{equation}\label{Amr3}
\begin{split}
A(2m+2,r;3)&=\frac{b(m,r-1)+b(m,r-2)}{2}\;A(2m+2;3)
\\
A(2m+3,r;3)&=\frac{2\,b(m,r-1)+5\,b(m,r-2)+2\,b(m,r-3)}{9}\;A(2m+3;3)
\end{split}
\end{equation}
where $b(m,\alpha)$ are defined as
\begin{equation}
B_{2m}(t)=t^mE_m(t)=:\sum_{\alpha=0}^{2m}b(m,\alpha)\,t^\alpha
\end{equation}
and  are assumed to be zero if $\alpha$ is out of the range of values
$0,1,\ldots,2m$. To find these coefficients we expand
\eqref{Efinal} in power series in $t$, thus expressing $B_{2m}(t)$
as a triple sum. Chu-Vandermonde summation formula can now be applied
to the sum with respect to the index defining the hypergeometric
series in \eqref{Efinal}, thus expressing $B_{2m}(t)$ as a double sum.
These two summations can be rearranged in such a way that one of them
becomes with respect to $\alpha$ while the other one defines the
coefficients of power expansion in $t$. We obtain
\begin{multline} \label{balpha}
b(m,\alpha)=\frac{(2m+1)!\,m!}{3^m\,(3m+2)!}
\sum_{\ell=\max(0,\alpha-m)}^{[\alpha/2]}
(2m+2-\alpha+2\ell)\binom{3m+3}{\alpha-2\ell}
\\ \times
\binom{2m+\ell-\alpha+1}{m+1} \binom{m+\ell+1}{m+1}2^{\alpha-2\ell}.
\end{multline}
Here $[\alpha/2]$ denotes integer part of $\alpha/2$.
Formulae \eqref{Amr3} and \eqref{balpha} complete the proof of the
Theorem, which constitutes the main result
of the present paper. Here we would like to mention that
in terms of terminating hypergeometric series the last formula
may be rewritten, for instance, as follows
\begin{multline}\label{bFF}
b(m,\alpha)=
\frac{2^{\alpha}\binom{3m+3}{\alpha} \binom{2m+1-\alpha}{m+1}}
{3^m\binom{3m+2}{m+1}}
\Biggl[
2 \Ffourthree{-(\alpha-1)/2}{-\alpha/2}
{m+2}{2m+2-\alpha}{(3m+4-\alpha)/2}{(3m+5-\alpha)/2}
{m-\alpha+1}{\frac{1}{4}}
\\
-\frac{\alpha}{m+1}
\Ffourthree{-(\alpha-1)/2}{-\alpha/2+1}
{m+2}{2m+2-\alpha}{(3m+4-\alpha)/2}{(3m+5-\alpha)/2}
{m-\alpha+1}{\frac{1}{4}}
\Biggr].
\end{multline}
This formula is valid for $\alpha=0,1,\ldots,m$
(a similar expression for $\alpha=m+1,m+2,\dots,2m$ can be simply
obtained through the replacement $\alpha\to 2m-\alpha$
in RHS of \eqref{bFF}). The two ${}_4F_3$ in \eqref{bFF} can be further
combined into a single ${}_5F_4\,$. Analogous formulae for $b(m,\alpha)$
in terms of terminating hypergeometric series of argument $4$ may
be written down as well.
Analyzing these expressions we were anyway unable to perform
the sum in \eqref{balpha} in a closed form, even if very suggestive
similarities
can be found with known summation formulae, see
\S\S 7.5 and 7.6, especially \S 7.6.4, of Ref.~\cite{PBM-III}.
We therefore regard
Eqns.~\eqref{Amr3} and \eqref{balpha} as the most
compact formulae for the numbers A(n,r;3) available at the present
moment.

To conclude, let us comment the
just obtained result
in application to the square ice. The quantity of interest
here is $nA(n,r;3)/A(n;3)$
which plays the role of spatial derivative
of the boundary polarization
of the square ice with a particular choice of the vertex weights.
It is interesting to study the behavior of this quantity in the
large $n$ and the large $r$ limits, with
the ratio $\xi=r/n$ kept fixed, $0<\xi<1$.
In the context of the square ice this
scaling limit corresponds to the continuous limit,
with $1/n$ playing the role of a lattice spacing.
For the square ice in the disordered regime
(which corresponds to $x$-enumerations
with $0<x<4$) a Heaviside step-function behavior $\theta(\xi-1/2)$
of the boundary polarization is expected \cite{BPZ-02}.
In ASM enumeration language this corresponds to
\begin{equation} \label{conjecture}
\lim_{\substack{n,r\to\infty\\ r/n=\xi}}n\,\frac{A(n,r;x)}{A(n;x)}
=\delta(\xi-1/2)
\qquad \text{for}\qquad 0<x<4.
\end{equation}
For instance, in the case of $1$-enumeration of ASMs (i.e., for $x=1$)
the validity of this formula can be verified directly from \eqref{Anr}
by virtue of Stirling formula. This result is also valid for
the simple case of $2$-enumerated ASMs whose refined
enumeration is just $A(n,r;2)/A(n;2)=\binom{n-1}{r-1}/2^{n-1}$;
for the result for the whole free-fermion line see \cite{BPZ-02}.
Formulae \eqref{Amr3} and \eqref{balpha} allows one to
study the scaling limit of the expression $nA(n,r;3)/A(n;3)$ as well.
In this limit the sum in \eqref{balpha} turns into an integral and
standard saddle point method can be applied.
In this way we find
\begin{equation}
\lim_{\substack{n,r\to\infty\\ r/n=\xi}}n\,\frac{A(n,r;3)}{A(n;3)}
=\delta(\xi-1/2).
\end{equation}
This confirms \eqref{conjecture} also in the  $x=3$ case.

\section*{Acknowledgments}

We would like to thank the referee for constructive comments.
We acknowledge financial support from MIUR COFIN programme  and from
INFN (Iniziativa Specifica FI11). One of us (A.G.P.) was also
supported in part by Russian Foundation for Basic Research, under
RFFI grant No. 04-01-00825, and by the programme
``Mathematical Methods in Nonlinear Dynamics'' of
Russian Academy of Sciences.
This work was been partially
done within the European Community
network EUCLID (HPRN-CT-2002-00325).


\end{document}